\documentclass[prb,preprint]{revtex4-1}
\usepackage{amsmath}  
\usepackage{amsfonts} 
\usepackage{graphicx} 
\begin{document}
\title{Radiation Reaction as a Non-conservative Force}
\author{Sandeep Aashish}
\email{ sandeepaashish@gmail.com} 
\author{Asrarul Haque }
\email{ahaque@hyderabad.bits-pilani.ac.in} \altaffiliation[A205,
Department of Physics]{BITS Pilani Hyderabad campus, Jawahar Nagar,
Shameerpet Mandal, Hyderabad-500078, AP, India.}
\affiliation{Department of Physics, BITS Pilani Hyderabad Campus,
 Hyderabad-500078, AP, India.}
\date{\today}
\begin{abstract}
We study a system of a finite size charged particle interacting with radiation field by exploiting the Hamilton's principle for non-conservative system introduced recently by Galley\cite{galley}. The said formulation leads to the equation of motion of the charged particle that turns out to be the same as obtained by Jackson\cite{jackson}. We show that
radiation reaction stems from the non-conservative piece of the effective Lagrangian.
We notice that 
 a charge interacting with radiation field modeled as heat bath affords a way to justify that radiation reaction is a non-conservative force. The topic is suitable for graduate courses on advanced electrodynamics and classical theory of fields.
\end{abstract}
\maketitle
\section{Introduction}
An accelerating charge radiates and therefore it loses energy\cite{jackson,schwinger,barut}. Larmor power accounts for the rate of loss of energy for a non-relativistic accelerated point charge. Loss of energy could be conceived to be caused by a damping force operating on the charge produced by the radiation field. This damping force is known as radiation reaction\cite{jackson}. \\
An accelerating charge is thus a non-conservative system which experiences radiation reaction - a non-conservative force. However, an accelerating charge together with its own field as a system conserves energy and is therefore conservative.\\
Although radiation reaction has been studied extensively in the past\cite{boyer,sharp,szeto,templin,templin2,fritz} 
with several notable attempts to rectify the issues of causality violation and runaway solutions\cite{cook,valentini,ford,connell,spohn,rohrlich2,griffiths}, it continues to be the subject-matter of several recent studies\cite{haque,haque2,steane,steane2}. 
We perform a systematic study of radiation reaction as a non-conservative force using Hamilton's principle for non-conservative systems to obtain the equation of motion of a charge and establish its non-conservative nature. \\
Radiation reaction has been studied in the light of Hamilton's principle in the past in an ad-hoc manner\cite{bateman,bender,barone} since the construction of the relevant Lagrangians in these papers are essentially the prescribed Lagrangians in Galley's formulation of Hamilton's principle. 
Polonyi\cite{polonyi} has studied radiation reaction problem as an effective theory using CTP(closed time path) formalism, however, his method has no immediate scope to construct a prescribed Lagrangian like Galley's formulation.\\
Galley\cite{galley} has formulated  Hamilton's principle compatible with initial data which leads to the Euler Lagrange equations of motion for both conservative as well as non-conservative systems. In the present work, we obtain the equation of motion of an accelerated charge by exploiting the Galley's formulation of Hamilton's principle for non-conservative systems. The equation of motion so obtained for a rigid, spherically symmetric charge instantaneously at rest turns out to be the same as obtained by Jackson\cite{jackson}. Radiation reaction is shown to stem from the non-conservative piece of the effective Lagrangian. 
The Abraham-Lorentz equation is derived using the effective non-conservative Lagrangian
for a point
charge. We notice on the ground of a correspondence 
between a charge interacting with radiation field and that of a particle interacting with infinite bath oscillators that radiation reaction could be realized as a non-conservative force.\\
In section 2, we observe that why the usual derivation of Euler-Lagrange equation from the effective action provides incomplete equation of motion in the case of an accelerated charge. In section 3, we derive the correct equation of motion for an accelerated charge using Galley's formulation of Hamilton's principle for non-conservative system.
\section{Preliminary}
We understand that Hamilton's principle is formulated as a boundary value problem in time to account for conservative systems. In this section, we shall briefly discuss the formulation of Hamilton's principle for non-conservative system due to Galley\cite{galley}.
\subsection{Hamilton's Action Principle for Non-conservative system }
Galley's formulation of Hamilton's principle compatible with initial value problems is endowed with a systematic algorithm as to how to obtain Euler-Lagrange equation for a non-conservative system. The essential ingredients required to obtain Euler-Lagrange equation for a non-conservative system are:
\begin{itemize}
\item[1.] a dynamical system and
\item[2.] coordinates $q_i$ and velocities ${\dot{q}_i}$.
\end{itemize}
 Consider a dynamical system with coordinates $q_i$ and velocities $\dot {q}_i$. 
 Action, a functional of the dynamical variables $q_i$, is defined by
\begin{equation}
S[q_i] = \int_{t_i}^{t_f}dtL(q_{i},\dot{q}_{i}).
\end{equation}
Under an arbitrary variation of the paths, $q_i(t)\rightarrow q_i(t) + \epsilon\eta_i(t)$, with the end points held fixed, i.e., $\eta(t_{i}) = 0 = \eta(t_{f})$, Hamilton's principle states that for the actual paths the first-order variation of $S$ vanishes
  \[\left(\frac{\delta S}{\delta \epsilon}\right)_{\epsilon=0}= {\left[ {\frac{\delta }{{\delta \varepsilon }}\int\limits_{{t_i}}^{{t_f}} {L({q_i}(t),{{\dot q}_i}(t))dt} } \right]_{\varepsilon  = 0}}
= 0\] 
which leads to the equation of motion:
  \[\frac{d}{dt}\frac{\partial L}{\partial \dot q_i(t)} -\frac{\partial L}{\partial q_i(t)} = 0;~ i = 1...n.\]
Hamilton's principle is formulated not as an initial value problem but as a boundary value problem in time since it requires the initial and final values of $q_i(t)$.
 However, the solutions to the derived equations of motion require initial data, i.e., $q_i(t)$ and $\dot {q}_i(t)$ at the initial time. It is this disparity between the two that lies at the heart of the incompatibility of the Hamilton's principle with dissipative (time-asymmetric) systems. To this end, we consider the following example.
Consider the action for a harmonic oscillator, 
\begin{eqnarray}
S = \int^{t_{f}}_{t_{i}} dt \left(\frac{1}{2}m\dot{x}^{2} - \frac{1}{2}kx^{2}\right)\label{2aex0}
\end{eqnarray}
which leads to the equation of motion, 
\begin{equation}
\left( {m\frac{{{d^2}}}{{d {t^2}}} + k} \right)x(t) = 0\label{2aex1}
\end{equation}
We can rewrite the action as:
\begin{eqnarray}
 \int\limits_{{t_i}}^{{t_f}} {\left( {\frac{1}{2}m{{\dot x}^2}(t) - \frac{1}{2}k{x^2}(t)} \right)dt}  &=& \int\limits_{{t_i}}^{{t_f}} 
 {\left( {\frac{1}{2}m\frac{d}{{dt}}(x\dot x) - \frac{1}{2}mx\ddot x - \frac{1}{2}k{x^2}(t)} \right)dt} \nonumber  \\ 
 & =& -\int\limits_{{t_i}}^{{t_f}} {\frac{1}{2}x(t)\left( { m\frac{{{d^2}}}{{d{t^2}}} + k} \right)x(t)dt} \label{2aex2}
 \end{eqnarray}
The total derivative term $\frac{1}{2}m\frac{d}{{dt}}(x\dot x)$ vanishes as the end points are held fixed. Note that the integrand in (\ref{2aex2}) is time-symmetric.
The equation of motion for the damped harmonic oscillator reads:
\[\left( {m\frac{{{d^2}}}{{d{t^2}}} + b\frac{d}{{dt}}}+k \right)x(t) = 0\]
Now, one might be tempted to write the action for the damped harmonic oscillator as
\[{S_d}[x(t)] =  - \int\limits_{{t_i}}^{{t_f}} {\frac{1}{2}x(t)\left( {m\frac{{{d^2}}}{{d{t^2}}} + b\frac{d}{{dt}} + k} \right)x(t)dt} \]
 Under the arbitrary variation of the path $x(t)\rightarrow x(t) + \epsilon\eta(t)$ with the endpoints fixed, $\eta(t_i)=0=\eta(t_f)$,
  the vanishing first-order variation of $S_d$
 \[
\left(\dfrac{\delta S_d}{\delta \epsilon}\right)_{\varepsilon  = 0} = 
-\frac{1}{2}\int^{t_{f}}_{t_{i}} dt~\left[(m\ddot{x} + kx)\eta + b\dfrac{d}{dt}(\eta x)\right] = 0
\]
 leads to the incorrect equation of motion
 \[\left( {m\frac{{{d^2}}}{{d{t^2}}} + k} \right)x(t) = 0\]
for the damped harmonic oscillator as the velocity term appears as total derivative and vanishes.\\
Galley has developed a consistent formulation of Hamilton's principle that is compatible with initial value problems.
 The formulation of Hamilton's principle for generic system is accomplished through 
the following steps:
\begin{itemize}
\item[Step I]Double both set of variables\cite{staru,schafer,faye}: $q_i \to (q_{i1},q_{i2})$ and $\dot{q}_i \to (\dot{q}_{i1},\dot{q}_{i2})$.
\begin{figure}
\centering
\includegraphics[scale=0.7]{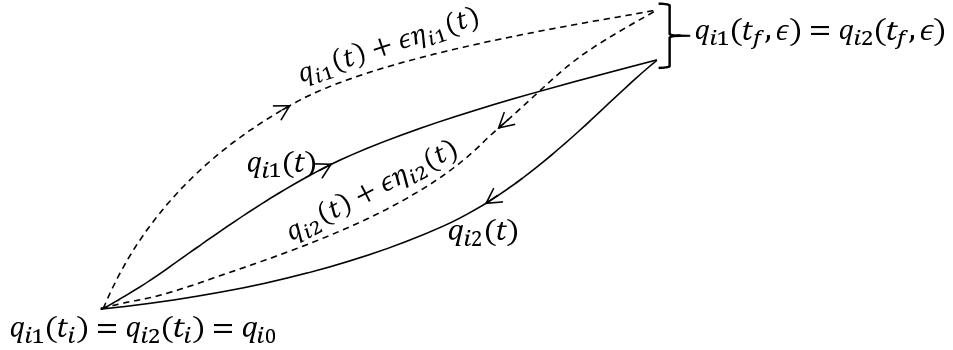} 
\label{FIG. 1}
\caption{The solid lines represent the paths $q_{i1}(t)$ and $q_{i2}(t)$ and the dashed lines stand for their infinitesimal ($\epsilon<<1$) departures. At the initial time $t_i$ both variables are fixed, i.e., $q_{i1}(t_i)= q_{i2}(t_i) = q_{i0}$ and at the final time $t_f$ their values coincide, i.e., $q_{i1}(t_f,\epsilon) = q_{i2}(t_f,\epsilon)$, but have an arbitrary variation.}
 \end{figure}
\item[Step II]
Define action, the functional of $q_{i1}$ and $q_{i2}$, as
\begin{equation}
S[q_{i1},q_{i2}] = \int_{ti}^{t_f}dt\Lambda(q_{i1,i2},\dot{q}_{i1,i2}) 
\end{equation}
where the new Lagrangian reads
\begin{equation}
\Lambda(q_{i1,i2},\dot{q}_{i1,i2}) = L(q_{i1},\dot{q}_{i1}) - L(q_{i2},\dot{q}_{i2})+ K(q_{i1,i2},\dot{q}_{i1,i2},t)\label{nl}
\end{equation}
and $K$ which depend on both set of variables encodes the important aspects of the formulation for
 when 
\begin{eqnarray*}
 K &=&0,~ \textup{two paths $q_{i1}$ and $q_{i2}$ are decoupled and $\Lambda$ represents a }\nonumber\\
 &&~~~~~\textup{conservative system,} \nonumber\\
 &\ne& 0,~\textup{ the two paths 
 $q_{i1}$ and $q_{i2}$ get coupled with each other and $\Lambda$ }\nonumber\\
 &&~~~~~\textup{represents a non-conservative system.} 
\end{eqnarray*}
It is thus the non-zero $K$ that determines the nonconsevative nature of the classical systems.
\item[Step III] Subject an arbitrary variation of the paths: $q_{i1,i2}(t)\to q_{i1,i2}(t,0)+\epsilon\eta_{i1,i2}(t)$
with $\eta_{i1,i2}(t_{i})=0$ at initial times and $q_{i1}(t_f,\epsilon)=q_{i2}(t_f,\epsilon)$ and $ \dot{q}_{i1}(t_f,\epsilon)=\dot{q}_{i2}(t_f,\epsilon)$(equality conditions) at final times so that Hamilton's principle for the corresponding change in the action becomes
\[{\left[ {\frac{{\delta S[{q_{i1}},{q_{i2}}]}}{{\delta \epsilon }}} \right]_{\scriptstyle \epsilon  = 0 \hfill \atop 
  \scriptstyle {q_{i1}} = {q_{i2}} = {q_i} \hfill}} = 0\]
where $q_{i1} = q_{i2} = q_i$ is called the \emph{physical limit}. Hamilton's principle leads to the following equation of motion
\[\frac{{\partial L}}{{\partial {q_i}}} - \frac{d}{{dt}}\frac{{\partial L}}{{\partial {{\dot q}_i}}} = {\left. {\left( {\frac{\partial }{{\partial {q_{i2}}}} - \frac{d}{{dt}}\frac{\partial }{{\partial {{\dot q}_{i2}}}}} \right)K({q_{i1,i2}},{{\dot q}_{i1,i2}},t)} \right|_{{q_{i1}} = {q_{i2}} = {q_i}}}\]
 It is noteworthy that the equality conditions imply that the data $q_{i1,i2}(t_{f})$ and $\dot q_{i1,i2}(t_{f})$ are not held fixed(see FIG.1.).
\item[Step IV] Change the variables (for convenience sake) $q_{i1,i2}(t)$ to:
\begin{equation}
q_{i+}\equiv \frac{q_{i1}+q_{i2}}{2}~~~~~\textup{and}~~~~~ q_{i-}\equiv q_{i1}-q_{i2}
\end{equation}
such that $q_{i-}\to 0$ and $q_{i+} \to q_i$ for $q_{i1}=q_{i2}$ (physical limit(PL)). The Hamilton's principle
\begin{equation}
{\left[ {\frac{{\delta S[{q_{i \pm }}]}}{{\delta \varepsilon }}} \right]_{\scriptstyle \varepsilon  = 0 \hfill \atop 
  \scriptstyle {q_i} = 0,{q_{i + }} = {q_i} \hfill}}
= \left[ \frac{\delta S[q_{i\pm}]}{\delta q_{i-}(t)}\right]_{PL}=0
\end{equation}
leads to the Euler-Lagrange equation of motion for general system as:
\begin{equation}
\frac{d{\pi}_i}{dt} = \left[ \frac{\partial\Lambda}{\partial q_{i-}}\right]_{PL} = \frac{\partial L}{\partial q_{i}} + 
\left[\frac{\partial K}{\partial q_{i-}}\right]_{PL} \label{ee}
\end{equation}
where the conjugate momenta ${\pi}_i$ are given by
\begin{equation}
\pi_i = \left[ \frac{\partial\Lambda}{\partial \dot{q}_{i-}}\right]_{PL} = \frac{\partial L}{\partial \dot{q}_{i}} + 
\left[\frac{\partial K}{\partial \dot{q}_{i-}}\right]_{PL}\label{mm} 
\end{equation}
\end{itemize}
$K$ is zero for conservative systems whereas open systems, which can exchange energy with the environment, possess non-zero $K$. A non-zero $K$ gives rise to non-conservative forces as we shall see in the later sections. Non-zero $K$ could be either
\begin{itemize}
\item \textit{prescribed} when the underlying variables that cause non-conservative (e.g. dissipative) forces are neither given nor modeled or
\item \textit{derived} when all the degrees of freedom of a closed system are given or modeled.
\end{itemize}
Non-zero prescribed $K$ could be obtained from the effective action (such as Fokker action) of the interacting systems. It is worthy to note that the interim variables ($q_{i\pm}$) need not be associated with any physical meaning until after the physical limit is applied. The novelty of the formulation lies in its generality in the sense that
\begin{itemize}
 \item it is compatible to both conservative as well as non-conservative systems and
 \item in this formulation it is possible to have a consistent Lagrangian consisting of distinct conservative and non-conservative part with the freedom to have either prescribed or derived $K$.
\end{itemize}

\section{\emph{Effective} Action For An Accelerated Charge}
Consider a system of finite-size charged particle coupled with its own electromagnetic field subjected to an external potential $V(\vec x)$. The Lagrangian of such a system of a charged particle plus electromagnetic field reads:
\begin{equation}
L = L_0 + L_{em} + L_{int}\label{ea7}
\end{equation} 
where,
\begin{eqnarray}
L_0
&=&\frac{1}{2}m\dot{x}_{i}^{2} - V(\vec x)\\
L_{em} & =&  \int d^3x\frac{1}{8\pi} \left( (-\vec{\nabla}\phi-\frac{1}{c}\dot{\vec{A}})^{2} - (\vec{\nabla}\times\vec{A})^{2}\right)       \nonumber\\
&=&\int d^3x\frac{1}{8\pi}\left((\partial_{i}\phi + \frac{1}{c}\dot{A}_{i})^{2} - (\partial_{i}A_{j}\partial_{i}A_{j}-\partial_{i}A_{j}\partial_{j}A_{i})\right)\\
L_{int}&=& \int d^3x (-\rho \phi + \frac{1}{c}J_iA_i)
\end{eqnarray}
$\rho$ and $J_i$ are charge and current densities of the charged particle respectively; $\phi$ and $A_i$ are the potentials associated to the self field of the charged particle. The action associated with the Lagrangian (\ref{ea7})
\begin{eqnarray}
S[{x_i},\phi ,{A_i}]&=& \int dt L({x_i},{{\dot x}_i},\phi ,\dot \phi ,{\partial _i}\phi ,{A_i},{{\dot A}_i},{\partial _j}{A_i})\nonumber\\
&=&\int{dt}\left(\frac{1}{2}m\dot{x}_{i}^{2} - V(\vec x)\right)\nonumber\\
&+&\int dtd^3x\frac{1}{8\pi}\left((\partial_{i}\phi + \frac{1}{c}\dot{A}_{i})^{2} - (\partial_{i}A_{j}\partial_{i}A_{j}-\partial_{i}A_{j}\partial_{j}A_{i})\right)\nonumber\\
&+& \int dt d^3x (-\rho \phi +\frac{1}{c} J_iA_i)
\end{eqnarray}
Variation of the action with respect to potentials $\phi$ and $A_i$ respectively in the Lorentz gauge yield the equations of motion for $\phi$ and $A_i$ as follows:
\begin{eqnarray}
\begin{array}{lcr}
\Box \phi &=& 4\pi \rho\\
\Box A_{i} &=& \dfrac{4\pi}{c}J_{i}
\label{ea6}
\end{array}
\end{eqnarray}
The solutions of the above equations are given by 
\begin{eqnarray}
\phi(\vec{x},t)& = &\int d^3x'dt' G_{ret}(\vec{x}-\vec{x}^{\prime}, t-t') \rho(\vec{x}',t)\label{ea5}\\
A_i(\vec{x},t)& =& \frac{1}{c}\int d^3x'dt' G_{ret}(\vec{x}-\vec{x}^{\prime}, t-t') J_i(\vec{x}',t)\label{ea4} 
\end{eqnarray}
Substituting (\ref{ea4}) and (\ref{ea5}) in (\ref{ea6}), we obtain the \emph{effective} action as,
\begin{eqnarray}
S_{eff}& =& \int dt \left[ \frac{1}{2}m\dot{x}_i^2 + \dfrac{1}{2c^2}\int d^3x\int d^{3}x' dt' \left\{J_i(\vec{x},t)J_i(\vec{x}^{\prime}, t') G_{ret}(\vec{x}-\vec{x}^{\prime}, t-t') \right.\right. \nonumber \\ 
&&\left.\left.  - ~c^2\rho (\vec{x}, t)\rho (\vec{x}^{\prime}, t') G_{ret}(\vec{x}-\vec{x}^{\prime}, t-t') \right\}\right]
\end{eqnarray}
The products $ J_i(\vec{x},t) J_i(\vec{x}^{\prime}, t')$ and $\rho (\vec{x}, t)\rho (\vec{x}^{\prime}, t')$ are symmetric under the exchange of variables ${x_i}\leftrightarrow x'_i$ and $t \leftrightarrow t'$ and couple only to the symmetric part of the retarded green function i.e.
\begin{eqnarray}
G_{+} &\equiv& \dfrac{G_{ret}(\vec{x}-\vec{x}^{\prime}, t-t') + G_{ret}(\vec{x}^{\prime}-\vec{x}, t'-t)}{2}\nonumber\\
& =&\dfrac{G_{ret}(\vec{x}-\vec{x}^{\prime}, t-t') + G_{adv}(\vec{x}-\vec{x}^{\prime}, t-t')}{2}
\end{eqnarray}
since $G_{ret}(\vec{x}^{\prime}-\vec{x}, t'-t) = G_{adv}(\vec{x}-\vec{x}^{\prime}, t-t')$. The effective action thus become 
\begin{eqnarray}
S_{eff} &=& \int dt \left[\frac{1}{2}m\dot{x}_{i}^{2} + \dfrac{1}{2c^2}\int d^3x\int d^{3}x' dt' \left\{ J_i(\vec{x},t)J_i(\vec{x}^{\prime}, t') G_{+}  \right.\right. \nonumber \\ 
&&\left.\left. -~ c^2\rho (\vec{x}, t)\rho (\vec{x}^{\prime}, t')G_{+} \right\}\right]\label {ea}
\end{eqnarray}
 The equation of motion of the finite-sized charged particle (rigid) could be obtained by applying Hamilton's principle to the effective action (\ref{ea}). We have 
\begin{eqnarray}
m\ddot{\vec x} + \int d^3x\int d^{3}x' dt'\rho(\vec{x},t)\left[\frac{1}{c^2}\dfrac{\partial G_{+}}{\partial t}\vec J(\vec{x}^{\prime},t') + \vec {\nabla}G_{+}\rho(\vec{x}^{\prime},t')\right] = 0\label{ea3}
\end{eqnarray}
where we have used for a rigid object, $J_i(\vec{x},t) = \rho(\vec{x},t){\dot{x}}_i(t)$. Equation (\ref{ea3}) would exhibit non-causal behavior for the presence of $ G_{adv}$ in $G_{+}$ would render the evolution of equation of motion acausal.
The second term of (\ref{ea3})\[\int d^3x\int d^{3}x' dt'\rho(\vec{x},t)\left[\frac{1}{c^2}\dfrac{\partial G_{+}}{\partial t}\vec J(\vec{x}^{\prime},t') + \vec {\nabla}G_{+}\rho(\vec{x}^{\prime},t')\right]\]subject to small $\frac{R}{c}$, rigid, spherically symmetric and instantaneously at rest charge distribution after performing a Taylor series expansion\cite{jackson} takes the form 
\[
\frac{1}{2}\sum_{n=0}^{\infty} \dfrac{2}{3}\left(\dfrac{1 + (-1)^n}{n!c^{n+2}}\right)\dfrac{\partial^{n+1}}{\partial t^{n+1}}\dot{\vec{x}}(t)\int d^3x\int d^{3}x' \rho (\vec x) R^{n-1} \rho(\vec{x}^{\prime})
\]
where $R=|\vec{x}-\vec{x'}|$. Equation (\ref{ea3})
now becomes 
\begin{eqnarray}
m\ddot{\vec{x}} + \frac{1}{2}\sum_{n=0}^{\infty} \dfrac{2}{3}\left(\dfrac{1 + (-1)^n}{n!c^{n+2}}\right)\dfrac{\partial^{n+1}}{\partial t^{n+1}}\dot{\vec{x}}(t)\int d^3x\int d^{3}x' \rho (\vec x) R^{n-1} \rho(\vec{x}^{\prime}) = 0\label{ea2}
\end{eqnarray}
which is not the correct equation of motion. Owing to merely temporally symmetric contribution to (\ref{ea2}), the second term in (\ref{ea2}) contains terms proportional to only even order time derivative like $\ddot{\vec{x}}(t),\ddddot{\vec{x}}(t)$ etc. and hence does not incorporate radiation reaction term which is proportional to $\dddot{\vec{x}}(t)$. It is the retarded green's function $G_{ret} = G_{+} + G_{-}$ that yields the correct equation of motion:
\begin{eqnarray}
m\ddot{\vec{x}} + \sum_{n=0}^{\infty} \dfrac{2}{3}\dfrac{(-1)^n}{n!c^{n+2}}\dfrac{\partial^{n+1}}{\partial t^{n+1}}\dot{\vec{x}}(t)\int d^3x\int d^{3}x' \rho (\vec x) R^{n-1} \rho(\vec{x}^{\prime}) = 0
\end{eqnarray} 
In the following section, we shall apply Galley's method\cite{galley} to a non-conservative system of an accelerated charge to obtain its correct equation of motion. 
\subsection{Equation of Motion for an Accelerated Charge}
The Lagrangian for an accelerated charge plus its electromagnetic field reads:
\begin{eqnarray}
L=\frac{1}{2}m\dot{x}_{i}^{2}-V(\vec x) + \int d^3x\left[\frac{1}{8\pi}\left((\partial_{i}\phi)^{2} + \frac{1}{c^2}\dot{A}_{i}^{2} + \frac{2}{c}(\partial_{i}\phi)\dot{A}_{i}- (\partial_{j}A_{i})^{2} + \partial_{j}A_{i}\partial_{i}A_{j}\right)\right. \nonumber \\ \left. -\rho\phi + \dfrac{1}{c}J_{i}A_{i}\right]
\end{eqnarray}
We begin with doubling the variables:
\begin{eqnarray}
x_{i} &\longrightarrow& x_{i1}, x_{i2} \nonumber \\
A_{i} &\longrightarrow& A_{i1}, A_{i2} \nonumber \\
\phi &\longrightarrow& \phi_{1}, \phi_{2}.
\end{eqnarray}
Now we subject the change of variables as given below:
\begin{eqnarray}
\begin{array}{lcr}
A_{i+}=\dfrac{A_{i1} + A_{i2}}{2} \\
A_{i-}= A_{i1}-A_{i2}
\end{array}
\begin{array}{lcr}
\phi_{+}=\dfrac{\phi_{1}+\phi_{2}}{2} \\
\phi_{-}=\phi_{1}-\phi_{2}
\end{array}
\begin{array}{lcr}
x_{i+}=\dfrac{x_{i1}+x_{i1}}{2} \\
x_{i-}=x_{i1}-x_{i2}
\end{array}
\end{eqnarray} 
so that the action becomes:
\begin{eqnarray}
S[x_{i+},x_{i-},\phi_{+},\phi_{-},A_{i+},A_{i-}] = \int dt \Lambda \nonumber\\
= \int dt \left[m\dot{x}_{i+}\dot{x}_{i-} - V_{-} + \int d^3x \left(\frac{1}{4\pi}\partial_{i}\phi_{+}\partial_{i}\phi_{-} + \frac{1}{4\pi c^2}\dot{A}_{i+}\dot{A}_{i-} \right.\right. \nonumber \\ \left.\left. -\frac{1}{4\pi}\partial_{j}A_{i-}\partial_{j}A_{i+} +\frac{1}{4\pi c}\dot{A}_{i-}\partial_{i}\phi_{+} + \frac{1}{4\pi c}\dot{A}_{i+}\partial\phi_{-} + \frac{1}{8\pi}\partial_{j}A_{i+}\partial_{i}A_{j-} \right.\right. \nonumber \\ \left.\left. + \frac{1}{8\pi}\partial_{j}A_{i-}\partial_{i}A_{j+} + \frac{1}{c}(J_{i+}A_{i-} + J_{i-}A_{i+})\right)\right]\label{dd}
\end{eqnarray}
where, $V_{-}\equiv V(\vec{x}_{1})-V(\vec{x}_{2})$. 
Equations of motion for $A_{i\pm}$ and $\phi_{\pm}$ are given by 
\begin{eqnarray}
\begin{array}{lcr}
\Box A_{i\pm} &=& \frac{4\pi}{c}J_{i\pm} \label{aa}\\
\Box\phi_{\pm} &=& 4\pi\rho_{\pm}
\end{array}
\end{eqnarray}
where we have identified,
\begin{eqnarray}
\rho_{+} &=& \dfrac{\rho_{1}+\rho_{2}}{2} = \frac{1}{2}\left[\rho(\vec{x}_{+} + \frac{1}{2}\vec{x}_{-}) + \rho(\vec{x}_{+} - \frac{1}{2}\vec{x}_{-})\right] \\
\rho_{-} &=& \rho_{1}-\rho_{2} = \rho(\vec{x}_{+} + \frac{1}{2}\vec{x}_{-}) - \rho(\vec{x}_{+} - \frac{1}{2}\vec{x}_{-})
\end{eqnarray} 
And 
\begin{eqnarray}
\begin{array}{lcr}
J_{i+} = \dfrac{J_{i1} + J_{i2}}{2} = \dot{x}_{i+}\rho_{+} + \frac{1}{4}\dot{x}_{i-}\rho_{-} \\ 
J_{i-} = J_{i1} - J_{i2} = \dot{x}_{i+}\rho_{-} + \dot{x}_{i-}\rho_{+}
\end{array}
\end{eqnarray}
We can solve the equations of motion for $A_{i\pm}$ in the Lorentz gauge. We assume that initial conditions are given for fields $\phi(t_i)$ and $A(t_i)$ as well as their first order spatial and temporal derivatives. These initial conditions could be associated with the initial conditions for $\phi_{+}$ and $A_{i+}$ in their physical limits.  
On the other hand, $\phi_{-}$ and $A_{i-}$ satisfy the equality condition at the final time given by
\begin{equation}
A_{i-}(t_{f}) = \phi_{-}(t_{f}) = \dot{\phi}_{-}(t_{f}) = \dot{A}_{i-}(t_{f})=0
\end{equation}
Now the solutions to (\ref{aa}) are
\begin{eqnarray}
{A}_{i+}(\vec{x},t) &=& \frac{1}{c}\int d^3x'dt' G_{ret}(\vec{x}-\vec{x}^{\prime}, t-t') \vec{J}_{+}(\vec{x}'_{\pm},t')\label{cc} \\
\phi_{+}(\vec{x},t) &=& \int d^3x'dt' G_{ret}(\vec{x}-\vec{x}^{\prime}, t-t') \rho_{+}(\vec{x}'_{\pm},t')\label{c1c}
\end{eqnarray}
upto solutions of the homogeneous equations, which have been set to zero and 
\begin{eqnarray}
A_{i-}(\vec{x},t) &=& \frac{1}{c}\int d^3x^{\prime} dt^{\prime} G_{adv}(\vec{x}-\vec{x}^{\prime}, t-t^{\prime})J_{i-}(\vec{x}^{\prime}_{\pm}, t^{\prime})\label{c2c} \\
\phi_{-}(\vec{x},t) &=& \int d^3x^{\prime} dt^{\prime} G_{adv}(\vec{x}-\vec{x}^{\prime}, t-t^{\prime})\rho_{-}(\vec{x}^{\prime}_{\pm}, t^{\prime})\label{c3c}
\end{eqnarray}
Substituting (\ref{cc}), (\ref{c1c}), (\ref{c2c}) and (\ref{c3c}) into the action (\ref{dd}), we obtain the effective action as 
\begin{eqnarray}
S_{eff} = \int dt \left[m\dot{x}_{i+}\dot{x}_{i-} - V_{-}  \right. \nonumber \\ \left. + \frac{1}{c^2}\int d^3x J_{i-}(\vec{x}_{\pm}, t)\int d^3x^{\prime} dt^{\prime} G_{ret}(\vec{x}-\vec{x}^{\prime}, t-t^{\prime})J_{i+}(\vec{x}^{\prime}_{\pm},t^{\prime})\right. \nonumber \\ \left. - \int d^3x \rho_{-}(\vec{x}_{\pm}, t)\int d^3x^{\prime} dt^{\prime} G_{ret}(\vec{x}-\vec{x}^{\prime}, t-t^{\prime})\rho_{+}(\vec{x}^{\prime}_{\pm},t^{\prime})\right]
\end{eqnarray}
which provides the effective Lagrangian
\begin{eqnarray}
\Lambda = m\dot{x}_{i+}\dot{x}_{i-} - V_{-} \nonumber \\ + \int d^3x d^3x^{\prime} dt^{\prime}\left[\frac{1}{c^2} J_{i-}(\vec{x}_{\pm}, t) G_{ret}(\vec{x}-\vec{x}^{\prime}, t-t^{\prime})J_{i+}(\vec{x}^{\prime}_{\pm},t^{\prime})\right. \nonumber \\ \left. - \rho_{-}(\vec{x}_{\pm}, t) G_{ret}(\vec{x}-\vec{x}^{\prime}, t-t^{\prime})\rho_{+}(\vec{x}^{\prime}_{\pm},t^{\prime})\right].\label{dr}
\end{eqnarray}
The non-conservative part of Lagrangian, $K$, can now be identified as
\begin{eqnarray}
K=\int d^3x d^3x^{\prime} dt^{\prime}\left[\frac{1}{c^2} J_{i-}(\vec{x}_{\pm}, t) G_{ret}(\vec{x}-\vec{x}^{\prime}, t-t^{\prime})J_{i+}(\vec{x}^{\prime}_{\pm},t^{\prime})\right. \nonumber \\ \left. - \rho_{-}(\vec{x}_{\pm}, t) G_{ret}(\vec{x}-\vec{x}^{\prime}, t-t^{\prime})\rho_{+}(\vec{x}^{\prime}_{\pm},t^{\prime})\right]
\end{eqnarray}
while the conservative part of the Lagrangian (in the physical limit) is
\begin{eqnarray}
L = \frac{1}{2}m\dot{x}_{i}^{2} - V(\vec{x})
\end{eqnarray}
Now Euler-Lagrange equation of motion 
\begin{equation}
\frac{d}{dt}\frac{\partial L}{\partial \dot {x}_{i}} + 
\frac{\partial L}{\partial{x}_{i}}=\left[ \frac{\partial K}{\partial x_{i-}} + 
\frac{d}{dt}\frac{\partial K}{\partial\dot{x}_{i-}}\right]_{PL}
\end{equation}
for $x_{i}(t)$ yields
\begin{eqnarray}
m\ddot{x}_{i} + \int d^3x\int d^{3}x' dt'\rho(\vec{x},t)\left[\frac{1}{c^2}\dfrac{\partial}{\partial t}[G_{ret}J_{i}(\vec{x}^{\prime},t')] \right. \nonumber \\ \left. + \partial_{i}[G_{ret}\rho(\vec{x}^{\prime},t')]\right] = F^{ext}_{i}\label{mmm}
\end{eqnarray}
where
\begin{eqnarray}
\left[\dfrac{\partial K}{\partial \dot{x}_{i-}}\right]_{PL}& =& \frac{1}{c^2}\int d^3x d^3x^{\prime} dt^{\prime}\rho(\vec{x},t) G_{ret}(\vec{x}-\vec{x}^{\prime}, t-t^{\prime})J_{i}(\vec{x}^{\prime},t^{\prime})
\\
\left[\dfrac{\partial K}{\partial x_{i-}}\right]_{PL} &=& - \int d^3x d^3x^{\prime} dt^{\prime}\rho(\vec{x},t) \partial_{i}[ G_{ret}(\vec{x}-\vec{x}^{\prime}, t-t^{\prime})\rho(\vec{x}^{\prime},t^{\prime})]\\
-\dfrac{\partial V}{\partial x_{i}}&\equiv& F^{ext}_{i}.
\end{eqnarray} 
 Moreover, for a rigid, spherically symmetric charged particle instantaneously at rest, equation of motion (\ref{mmm}) takes the form:
\begin{eqnarray}
m\ddot{\vec{x}} + \sum_{n=0}^{\infty} \dfrac{2}{3}\dfrac{(-1)^n}{n!c^{n+2}}\dfrac{\partial^{n+1}}{\partial t^{n+1}}\dot{\vec{x}}(t)\int d^3x\int d^{3}x' \rho (\vec{x}) R^{n-1} \rho(\vec{x}^{\prime}) = \vec{F}_{ext}\label{qqq}
\end{eqnarray}
This is in fact the same equation as obtained by Jackson \cite{jackson}. 
We observe that the radiation reaction in (\ref{qqq}) is indeed derivable from $K$. This is a remarkable result, since it is now possible to distinctly obtain non-conservative forces from non-conservative part ($K$) of the effective Lagrangian.
At this juncture, it is feasible now to prescribe a Lagrangian that would result the same equation of motion as (\ref{qqq}): 
\begin{eqnarray}
\Lambda = m\dot{\vec{x}}_{+}\cdot\dot{\vec{x}}_{-} + \vec{x}_{-}\cdot\vec{F}_{ext} - \sum_{n=0}^{\infty} \dfrac{2}{3}\dfrac{(-1)^n}{n!c^{n+2}}\vec{x}_{-}\cdot\dfrac{\partial^{n+1}}{\partial t^{n+1}}\dot{\vec{x}}_{+}I_{n}\label{pr}
\end{eqnarray}
where, 
\begin{eqnarray}
I_{n}\equiv \int d^3x\int d^{3}x' \rho (\vec{x}) R^{n-1} \rho(\vec{x}^{\prime})
\end{eqnarray}
We note that the prescribed Lagrangian (\ref{pr}) is seemingly different from the derived one (\ref{dr}), however, both of them lead
to the same equation of motion in the physical limit. We can observe as an advantage of this formalism that we have the freedom to choose $K$ (the last term in (\ref{pr})). For example, $K= +\sum_{n=0}^{\infty} \dfrac{2}{3}\dfrac{(-1)^n}{n!c^{n+2}}\dot{\vec{x}}_{-}\cdot\dfrac{\partial^{n}}{\partial t^{n}}\dot{\vec{x}}_{+}I_{n}$ is an equally acceptable choice.
\subsection{Equation of motion for a point charge from effective Lagrangian}
\sectionmark{Physically correct equation...}
The effective Lagrangian for a point charge can be obtained from $\Lambda$ in (\ref{dr}), by making use of the Green's function
\begin{eqnarray}
G_{ret}(\vec{x}-\vec{x}^{\prime}, t-t^{\prime}) = \frac{1}{R}\delta(t' - t + \frac{R}{c})
\end{eqnarray}
and integrating over $t'$, as follows:
\begin{eqnarray}
\Lambda = m\dot{z}_{i+}\dot{z}_{i-} - V_{-} + \frac{1}{c^2}\int d^3x d^3x' J_{i-}(\vec{x},\vec{z}_{\pm},t)\frac{1}{R}J_{i+}(\vec{x}^{\prime},\vec{z}_{\pm},t_{r}) \nonumber \\ - \int d^3x d^3x'\rho_{-}(\vec{x},\vec{z}_{\pm},t) \frac{1}{R}\rho_{+}(\vec{x}^{\prime},\vec{z}_{\pm},t_{r}) 
\end{eqnarray}
where $t_{r} = t - R/c$ and $V_{-}$ is given by 
\begin{eqnarray}
V_{-} = V(\vec{z}_{+}+\frac{1}{2}\vec{z}_{-}) - V(\vec{z}_{+}-\frac{1}{2}\vec{z}_{-})
\end{eqnarray}
If the velocity of charge is small compared to that of light, then the change in charge and current densities could be considered negligible over the time interval $R/c$ \cite{landau}. Hence, in the point charge limit ($R\to 0$), we can neglect the higher order terms in the expansion of $\rho_{+}(t_{r})$ and $J_{i+}(t_{r})$ about $t$: 
\begin{eqnarray}
J_{i+}(t_{r}) &\approx & J_{i+}(t) - \dfrac{R}{c}\dot{J}_{i+}(t) \\
\rho_{+}(t_{r}) &\approx & \rho_{+}(t) - \frac{R}{c}\dot{\rho}_{+}(t) + \dfrac{R^2}{2c^2}\ddot{\rho}_{+}(t) - \dfrac{R^3}{6c^3}\dddot{\rho}_{+}(t) 
\end{eqnarray}
Using the above expansions, the effective Lagrangian, $\Lambda$, can now be written as,
\begin{eqnarray}
\Lambda = m\dot{z}_{i+}\dot{z}_{i-} - V_{-} + \nonumber \\ \int d^3x d^3x' \left[\frac{1}{c^2}J_{i-}(\vec{x},\vec{z}_{\pm},t)\frac{1}{R}J_{i+}(\vec{x}^{\prime},\vec{z}_{\pm},t) - \frac{1}{2c^2}\rho_{-}(\vec{x},\vec{z}_{\pm},t)R\ddot{\rho}_{+}(\vec{x}^{\prime},\vec{z}_{\pm},t) \right. \nonumber \\ \left. - \frac{1}{c^3}J_{i-}(\vec{x},\vec{z}_{\pm},t)\dot{J}_{i+}(\vec{x}^{\prime},\vec{z}_{\pm},t) + \frac{1}{6c^3}\rho_{-}(\vec{x},\vec{z}_{\pm},t)R^{2}\dddot{\rho}_{+}(\vec{x}^{\prime},\vec{z}_{\pm},t) \right. \nonumber \\ \left. - \rho_{-}(\vec{x},\vec{z}_{\pm},t)\frac{1}{R}\rho_{+}(\vec{x}^{\prime},\vec{z}_{\pm},t) + \frac{1}{c}\rho_{-}(\vec{x},\vec{z}_{\pm},t)\dot{\rho}_{+}(\vec{x}^{\prime},\vec{z}_{\pm},t)\right]
\end{eqnarray}
The equation of motion 
\begin{eqnarray}
\dfrac{d}{dt}\left[\dfrac{\partial\Lambda}{\partial \dot{z}_{i-}}\right]_{PL} = \left[\dfrac{\partial\Lambda}{\partial z_{i-}}\right]_{PL}
\end{eqnarray}
of the point charge yields
\begin{eqnarray}
m\ddot{z}_{i} + \frac{1}{c^2}\int d^3x d^3x' \rho(\vec{x}-\vec{z},t)\frac{1}{R}\dfrac{\partial}{\partial t}\left[J_{i}(\vec{x}^{\prime},t) + \dfrac{\partial_{i} R}{2R^{-1}}\dot{\rho}(\vec{x}^{\prime}-\vec{z},t)\right] \nonumber \\ - \frac{1}{c^3}\int d^3x d^3x' \rho(\vec{x}-\vec{z},t)\dfrac{\partial^2}{\partial t^2}\left[J_{i}(\vec{x}^{\prime},t) + \dfrac{\partial_{i}R^{2}}{6}\dot{\rho}(\vec{x}^{\prime}-\vec{z},t)\right] = F^{ext}_{i}\label{ssd0}
\end{eqnarray}
where, 
\begin{eqnarray}
\left[\dfrac{\partial\Lambda}{\partial \dot{z}_{i-}}\right]_{PL} &=& m\dot{z}_{i} + \int d^3x d^3x' \rho(\vec{x}-\vec{z},t)\left[\frac{1}{c^2}\frac{1}{R}J_{i}(\vec{x}^{\prime},\vec{z},t) - \frac{1}{c^3}\dot{J}_{i}(\vec{x}^{\prime},\vec{z},t)\right] \nonumber
\\
\left[\dfrac{\partial\Lambda}{\partial z_{i-}}\right]_{PL} &=& -\dfrac{\partial V}{\partial x_{i}} - \int d^3x d^3x' \left[\frac{1}{2c^2}\rho(\vec{x}-\vec{z},t)\partial_{i}R\ddot{\rho}(\vec{x}^{\prime}-\vec{z},t) \right. \nonumber \\&& \left.- \frac{1}{6c^3}\rho(\vec{x}-\vec{z},t)\partial_{i}R^{2}\dddot{\rho}(\vec{x}^{\prime}-\vec{z},t)+ \rho(\vec{x}-\vec{z},t)\partial_{i}\frac{1}{R}\rho(\vec{x}^{\prime}-\vec{z},t) \right. \nonumber \\&& \left.  - \frac{1}{c}\partial_{i}\rho(\vec{x}-\vec{z},t)\dot{\rho}(\vec{x}^{\prime}-\vec{z},t)\right]\nonumber\\
0 &=& \int d^3x d^3x' \rho(\vec{x}-\vec{z},t)\partial_{i}\left(\dfrac{1}{R}\right) \rho(\vec{x}'-\vec{z},t)\nonumber \\
0&=& \partial_{i}\int d^3x d^3x' \rho(\vec{x}-\vec{z},t)\dot{\rho}(\vec{x}'-\vec{z},t) \nonumber
\end{eqnarray}
For a spherically symmetric charge distribution (see p. 752-753 of ref. \cite{jackson}), 
\begin{eqnarray}
\vec{J}(\vec{x}^{\prime},t) + \dfrac{\partial\rho}{\partial t}(\vec{x}^{\prime}-\vec{z},t)\dfrac{\partial_{i} R^{n+1}}{(n+1)(n+2)R^{n-1}} = \frac{2}{3}\dot{z}_{i}\rho(\vec{x}^{\prime}-\vec{z},t)\label{ssd}.
\end{eqnarray} 
Using the above result in (\ref{ssd0}) and substituting $\rho(\vec{x}-\vec{z})=e\delta(\vec{x}-\vec{z})$ and $J_{i}(\vec{x},t)=e\dot{z}_{i}\delta(\vec{x}-\vec{z})$, followed by integrating over $x$ and $x'$, we obtain the equation of motion of a point charge as
\begin{eqnarray}
m\ddot{\vec{z}} + \lim_{R\to 0}\dfrac{2e^2}{3c^2R}\ddot{\vec{z}} - \dfrac{2e^2}{3c^3}\dddot{\vec{z}} = \vec{F}_{ext}.
\end{eqnarray}
The term $\dfrac{2e^2}{3c^2R}$ which is divergent in the limit $R \to 0$ possesses the dimension of mass and couples with the bare mass $m$ to give the renormalized mass M,
\begin{eqnarray}
M = m+\lim_{R\to 0}\dfrac{2e^2}{3c^2R}
\end{eqnarray} 
so that the equation of motion can now be expressed as the famous Abraham-Lorentz equation,
\begin{eqnarray}
M(\ddot{\vec{z}}-\tau\dddot{\vec{z}}) = \vec{F}_{ext}\label{abl}
\end{eqnarray}
We note that the Abraham-Lorentz equation can also be obtained directly from (\ref{qqq}) by substituting $\rho(\vec{x})= e\delta(\vec{x})$ and taking $R\to 0$. \\
We know that the Abraham-Lorentz equation suffers from the problems of causality-violation and runaway solutions. It has been shown by Rohrlich\cite{rohrlich2, rohrlich3} that if the external force varies slowly over the size of a charged particle (of the order $c\tau$) such that
\begin{eqnarray}
\left|\tau\dfrac{d}{dt}\vec{F}_{ext}\right| \ll |\vec{F}_{ext}(t)|
\end{eqnarray}
then the (finite) charge distribution can be approximated as a point charge and the effect of radiation reaction on the motion of charge is negligible. This allows us to write 
\begin{equation}
M\ddot{\vec{z}} \approx \vec{F}_{ext}
\end{equation}
or, 
\begin{eqnarray}
\dddot{\vec{z}} \cong \dfrac{\dot{\vec{F}}_{ext}}{M}.
\end{eqnarray}
The equation of motion obtained by substituting for $\dddot{\vec{z}}$ in (\ref{abl}) is free of the above-mentioned issues: causality-violation and runaway solutions and is given by
\begin{eqnarray}
M\ddot{\vec{z}} = \vec{F}_{ext} + \tau\dot{\vec{F}}_{ext}
\end{eqnarray} 
A Lagrangian may now be prescribed for a system of an accelerated charge in the light of the above equation of motion as follows
\begin{equation}
\Lambda = M\dot{\vec{z}}_{+}\cdot\dot{\vec{z}}_{-} + \vec{z}_{-}\cdot\vec{F}_{ext} - \tau\dot{\vec{z}}_{-}\cdot\vec{F}_{ext}.
\end{equation}
Moreover the non-conservative part of the Lagrangian reads 
\begin{eqnarray}
K = \vec{z}_{-}\cdot\vec{F}_{ext} - \tau\dot{\vec{z}}_{-}\cdot\vec{F}_{ext}.
\end{eqnarray}
\section{Conclusion and Discussion}
We have shown that Hamilton's principle for non-conservative system such as a finite size charged particle interacting with radiation field furnishes the same equation of motion as that obtained by Jackson\cite{jackson}.
Moreover, in the limit of a point charge, we obtain the Abraham-Lorentz equation.  We find that radiation reaction is derivable from the non-conservative part of the effective Lagrangian.  
A systematic study of radiation reaction, based on the interaction of a finite size charged particle with radiation field modeled as a heat bath, has been made in the past \cite{ford,ford2}. This model affords a way to justify that an accelerating charge interacting with radiation field is a non-conservative system. Since, for a particle interacting with a heat bath modeled as a large number of independent harmonic oscillators, if the number of the bath oscillators is about $20$ or larger, the Poincar$\acute{\textnormal{e}}$ recurrence time (amount of time after which a system returns to a state very close to the initial state) turns out infinite\cite{weiss,grie} which renders the dynamics of interacting particle non-conservative.

\end{document}